\documentclass[prb,twocolumn,amsmath,amssymb,floatfix,footinbib,bibnotes]{revtex4-1}

\usepackage[colorlinks=true,citecolor=blue,linkcolor=magenta]{hyperref}
\usepackage{amsmath}
\usepackage{graphicx}
\usepackage{dsfont}
\usepackage{changepage}
\usepackage{fancyhdr}
\usepackage{amsthm, amssymb}
\usepackage{array}
\usepackage[usenames,dvipsnames]{color}

\newcommand{\be}{\begin{align}}
\newcommand{\ee}{\end{align}}

\def \be{\begin{equation}}
\def \ee{\end{equation}}
\def \ba{\begin{array}}
\def \ea{\end{array}}
\def \bea{\begin{eqnarray}}
\def \eea{\end{eqnarray}}
\def \nn{\nonumber}
\def \W{{\Omega}}
\def \e{{\epsilon}}
\def \l{{\lambda}}

\def \a{{\alpha}}

\def \d{{\delta}}
\def \w{{\omega}}

\def \f{{\varphi}}

\def \e{{\epsilon}}
\def \ve{{\varepsilon}}

\def \ba{\begin{align*}}
\def \ea{\end{align*}}

\newcounter{indice}

\def \mrm{\mathrm}

\def \bs{\boldsymbol}
\def \mc{\mathcal}

\begin{document}

\title{Comment on "Superconductivity at low density near a ferroelectric quantum critical point: Doped
SrTiO$_3$"    }

\author{Jonathan Ruhman$^1$ and Patrick A. Lee$^2$}
\affiliation{ {
1. Department of Physics, Bar Ilan University, Ramat Gan 5290002, Israel \\
2. Department of Physics, Massachusetts Institute of Technology, Cambridge, MA 02139 USA  }}
\begin{abstract}
{
W\"olfle and Balatsky Phys. Rev. B 98, 104505 (2018) have proposed a microscopic pairing mechanism for doped SrTiO$_3$ (STO) based on the {\it gradient}
coupling of electronic density to the soft TO phonon mode. Since this coupling to TO phonons is usually weak, this conclusion is surprising, especially for a
low density superconductor such as STO, where the density of states is small.  A crucial step in the argument made by W\"olfle and Balatsky is that the
displacement vector of the TO mode is not strictly perpendicular to the momentum vector, making a deformation coupling possible. We show that they have made a
mistake in computing the eigenvector and have grossly overestimated this lack of orthogonality.
When corrected, the coupling is negligible.
We also use transport data to put upper bounds on the coupling constant which are much smaller than the estimate by W\"olfle and Balatsky.  Finally, we also
object to their use of the Eliashberg equation when the phonon frequency is larger than the Fermi energy.
 }
\end{abstract}

\maketitle

The microscopic mechanism for superconductivity in doped SrTiO$_3$ (STO) has been recently discussed in Ref.~\onlinecite{Wolfle2018}.\footnote{ The objective of Ref.~\onlinecite{Wolfle2018} was to fill in a gap formed in Ref.~\onlinecite{Edge2015}, where it was argued that the paraelectric-ferroelectric quantum critical point in STO\cite{Rowley2014} is the origin of the superconducting dome without specifying the microscopic coupling between the Fermi surface and the soft TO mode.  } Central to their discussion is the claim that exchanging the soft transverse optical (TO) mode, that is related to the proximity to the ferroelectric transition,
gives a sufficiently large coupling to explain the experimentally observed $T_c$, even at very low densities. This is contrary to the  statement
we made in an earlier publication~\cite{Ruhman2016} that due to the low density of states, the dimensionless electron phonon coupling $\l$ via the exchange of TO
phonons is too weak to give any reasonable $T_c$. In addition, we also stated that while the mechanism of exchanging polar phonon proposed by Gurevich {\it et.
al.}~\cite{gurevich} may work at intermediate doping densities, even that fails for very low densities when the requirement for adiabaticity of the phonons is
respected.
The work of Ref.~\onlinecite{Wolfle2018} motivated us to study in detail the coupling of TO phonons to electrons near the zone center. We clarified the roles of
Coulomb interaction with the ions versus that of a short range pseudo-potential. In the process we found an error made in Ref.~\onlinecite{Wolfle2018}, which led
them to overestimate the coupling strength and we re-affirm our previous conclusion.

In the case of a ferroelectric transition the order parameter is a lattice distortion, i.e. an optical phonon mode. There are three relevant phonon
polarizations, which are traditionally labeled as one longitudinal optical (LO) and two transverse optical (TO) modes. The long-ranged dipolar interactions in
the LO mode, however, make it stiff and prevent it from becoming soft at the transition~\cite{khmelnit1971low,strukov2012ferroelectric,kvyatkovskii2001quantum,roussev2003theory}. Thus, the two TO modes are the dynamical soft
boson modes associated with this transition.

To get pairing directly from these soft modes the authors of Ref.~\onlinecite{Wolfle2018} used a phenomenological model involving a displacement vector $\bs
u_{\bs q}$ and invoked a gradient coupling of the form
\be
H_{el-ph}=iV_0 \sum_{\bs q}\bs q\cdot  \bs u_{\bs q} \rho_{-\bs q} \label{el-ph}
\ee
where $\bs u$ is the soft phonon displacement and $\rho$ is the electronic density. Clearly, if the polarizations of the TO modes are truly transverse this
coupling vanishes. They argued that cubic crystal anisotropy tilts the polarization of the soft modes such that they are never really transverse to
$\hat{\bs q}$ except for high symmetry lines. The square of the overlap between the mode polarization $\hat{\bs e}_T(\bs q)$ and the momentum direction
\be
s\equiv\langle (\hat{\bs e}_T(\bs q)\cdot\hat{\bs q})^2\rangle  \label{s}
\ee
was estimated and found to be of order 0.1.
This parameter multiplies the effective BCS coupling in the s-wave channel, which arises from Eq.~\eqref{el-ph}. Thus, their theory depends crucially on the fact
that $s$ is order 0.1 and not negligibly small.

It is important to note that for STO the carrier density and therefore the Fermi momentum is very small and the small $q$ limit of Eq.~\eqref{el-ph} is relevant.
Physically we expect the crystalline symmetry to become irrelevant in the $q \rightarrow 0$ limit and the polarization should be strictly perpendicular and
parallel to $\bs q$ for the TO and LO mode, respectively. As a result $s$ should approach zero. This is because any deviation from strictly transverse
displacement will mix in long range polarization and raise the energy of the TO mode. In the following we prove this analytically and obtain an expression
showing that $s$ is proportional to $q^4$ (see Eq. \eqref{s_correct} below). Setting $q$ to be the $k_F$ for a given density, we find that for a density $n = 10^{20}
\,\mrm{cm^{-3}}$, $s$ is smaller than $10^{-5}$ and drops drastically with density.
Thus, the authors of Ref.~\onlinecite{Wolfle2018} have miscalculated $s$ by orders of
magnitude. Their estimate of $\l$ is off by the same amount and their conclusion that $T_c$ in STO can be explained by coupling to the soft TO phonon should be reevaluated.

In what follows, we also pin-point the error made in Ref.~\onlinecite{Wolfle2018}. In evaluating the TO mode eigenvectors, they did not take into account the
long-ranged Coulomb forces, which are dominant in the limit $q \rightarrow 0$.  Consequently, they find $s$ to be independent on  $q$ and substantial, as in
non-polar materials.

We then go beyond the phenomenological model and consider the full problem of a crystal structure with multiple ions, as is appropriate for STO. We find that the
results of the phenomenological model applies to the Coulomb potential considered by W\"olfle and Balatsky, which gives a negligible contribution to the
coupling. There exists a second kind of contribution from the short range pseudo-potential, which is not generally forced to vanish in the same way. However, for
cubic and tetragonal crystals it takes the form Eq.~\eqref{el-ph} and  is also negligible due to the same factor.

Another issue in Ref.~\onlinecite{Wolfle2018} regards the other pairing mechanism they have considered: the high frequency LO mode. The authors continued to use the Eliashberg
equations even when the phonon frequency exceeds the Fermi energy by one and two orders of magnitude. We argue at the end of this comment that in the  absence of
proof to the contrary, conventional wisdom should apply and the Eliashberg equation should be restricted to the adiabatic limit.

\section{optical-phonon Dispersion close to the zone center in a polar cubic crystal}
In this section we present the action for optical phonons in SrTiO$_3$ from which the dispersion and polarization vectors near the zone center can be derived.
Because the error in Ref.~\onlinecite{Wolfle2018} arises from ambiguity in the inclusion of long-ranged forces in the action formalism we first include
dispersion terms  based on short-ranged forces, which are identical to the ones in Ref.~\onlinecite{Wolfle2018}. Then, in the next step we show how to include
the long-ranged forces.

We start from the action
\be
\mc S_u  = {1\over 2}\sum_{\w, \bs q}{u_{-\bs q}^ j \left[ \w^2 \d_{jl}- A_{jl}(\bs q) \right]{u_{\bs q}^ l}}\label{S0}
\ee
Here $\bs u$ is the optical displacement which is proportional to the dipolar polarization and $A_{ij}(\bs q)$ is the dispersion matrix given by
\be
A_{ij}(\bs q)=\w_T^2 \d_{ij} + c_T^2  \left(q^2\d_{ij}- q_i  q_j \right)+c_L^2 q_i q_j  +\a q_i^2 \d_{ij}
\ee
$c_L$ and $c_T$ are the longitudinal and transverse phonon velocities, respectively, $\w_T$ is the optical gap at the zone center and $\a$ arises from the cubic
crystal fields, which breaks rotational symmetry. The model above is identical to the one used by W\"olfle and Balatsky~\cite{Wolfle2018} (see Eq. A15). The two
are related using the identities $\bs P = \w_D \bs u$, $\w_D = 33$ meV, $\l_1 =  \left(c_L^2 + \a\right)/\w_D^2 = 8 \,\mrm\AA^2$, $\l_2 = \left(
c_L^2-c_T^2\right)/\w_D^2 = 1 \,\mrm\AA^2$, $\l_3 = c_T^2/\w_D^2 2 \,\mrm\AA^2 $ and $\tau =  \w_T^2/\w_D^2$. In the first identity $\bs P$ is essentially the dipolar
polarization associated with the optical distortion field $\bs u$. Note that the numbers quoted here are copied from Ref.~\onlinecite{Wolfle2018}, and we have not
independently verified their accuracy.

We also note that we insist on modifying the notations of Ref.~\onlinecite{Wolfle2018} to the ones in Eq.~\eqref{S0} because in this representation the role of
the crystal anisotropy $\a$ becomes clear. This also allows us to seperate the terms which are fully rotationally symmetric form this term, which will become
useful in the next section.

To add the effects of the long-ranged dipolar forces associated with the optical distortion we write their coupling to the electric field
\be
\mc S_{E-u} = -\sum_{\bs q , \w} \bs P_{\bs q}\cdot \bs E_{-\bs q}=-\w_D\sum_{\bs q , \w} \bs u_{\bs q}\cdot \bs E_{-\bs q}\label{S_Eu1}
\ee
It is crucial to note that we assume that $\bs E$ is static (we neglect dynamics of the electromagnetic fields). Thus, $\bs E$ stems from a potential $\f$, such
that $\bs E_{\bs q} = i\bs q \cdot \f_{\bs q}$. Consequently, $\bs E_{\bs q}$ is strictly a longitudinal vector. When taking the dot-product with $\bs u_{\bs q}$
in the coupling term \eqref{S_Eu1}, it nullifies the components perpendicular to $\bs q$ and we can equivalently write the coupling term as
\be
\mc S_{E-u} = -\w_D\sum_{\bs q , \w} \hat{\bs q}\cdot\bs u_{\bs q} E_{-\bs q}\label{S_Eu}
\ee
To obtain the effective dispersion we would like to integrate out the electric field. To this end, we recall the action of the of the electric field, which is
the energy density (again, neglecting dynamics)
\be
\mc S_E ={\ve _{\infty} \over 8\pi}\sum_{\bs q,\w} { | E_{\bs q}|^2  } \label{S_E}
\ee
Completing a square between Eq.~\eqref{S_E} and Eq.~\eqref{S_Eu} and integrating over the electric field we obtain a shift in the field $\bs u$
\be \d \mc S_{u} = -{1\over 2}\sum_{\w\bs q}{ 4\pi \w_D^2 \over \ve_{\infty}  } \hat q_j \hat q_l u_{\bs q}^j u_{-\bs q}^l \label{shift}\ee
Thus, the combined effect of long and short ranged physics leads to the action
\be
\tilde{\mc S}_u=\mc S_u+\d \mc S_u  = {1\over 2}\sum_{\w, \bs q}{u_{-\bs q}^ j \left[ \w^2 \d_{jl}- \tilde A_{jl}(\bs q) \right]{u_{\bs q}^ l}}\label{S0_correct}
\ee
where the corrected dispersion matrix is given by~\cite{khmelnit1971low}
\begin{align}
\tilde A_{ij}(\bs q)=\w_T^2 \d_{ij}& + c_T^2  \left(q^2\d_{ij}- q_i  q_j \right)\label{Atilde}\\&+\left( \w_L^2 - \w_T^2 \right)  \hat q_i \hat q_j+c_L^2 q_i q_j
+\a q_i^2 \d_{ij}\nn
\end{align}
The  LO and TO modes are the eigenstates of this equation. Note that as expected, the LO frequecy has been sifted from $\w_T$ to $\w_L \equiv \sqrt{\w_T^2 + {
4\pi \w_D^2 / \ve_{\infty}  }}$.

In SrTiO$_3$ this mass term is quite large. Neglecting complications coming from multiple modes, we can assign $\w_L \approx 100 \, \mrm{meV}$, while $\w_T
\approx 2 \, \mrm{meV}$. The quantum critical point is obtained by taking $\w_T \rightarrow 0$. The dynamics of the longitudinal component, however, are
described by a finite frequency mode.

\section{Estimation of the polarization vectors and dispersion}\label{Sec:polariaztions}
The polarization of the optical phonon branches are dictated by the dispersion matrix $\tilde A(\bs q)$.
\textbf{The main mistake of the authors of Ref.~\onlinecite{Wolfle2018} is that they computed the soft mode polarization vectors $\hat{\bs e}_T$ using the matrix
$A(\bs q)$ in action \eqref{S0} and not $\tilde A (\bs q)$ in the action Eq.~\eqref{S0_correct} (see Eq.(A30) in their appendix)}. As a result they obtain that
the polarization vectors are independent on the magnitude of the momentum $q$. However, upon inspection of Eq.\eqref{S0_correct} we find that in the limit of
$q\rightarrow 0$ the mass term Eq. \eqref{shift} remains finite, unlike the anisotropy term $\a$. In this limit the gapped mode is purely longitudinal and
decoupled from the soft transverse modes, thus nullifying the coupling shown in Eq.~\eqref{el-ph} to the TO modes. This conclusion holds for any TO mode that
involves a dipolar excitation, whether it is soft or not.

To see this let us estimate the polarization vectors close to $q=0$.
We first separate the dispersion matrix into a fully rotational symmetric part and the anisotropy term
\be
\tilde A(\bs q) = M(\bs q) + \d M(\bs q)
\ee
Here $M_{ij}(\bs q) =\w_T^2 \d_{ij} + c_T^2  \left(q^2\d_{ij}- q_i  q_j \right)+(\w_L^2 - \w_T^2){q_i q_j \over q^2}+c_L^2 q_i q_j $ is the isotropic matrix and
$\d M_{ij} (\bs q) = \a q_i^2 \d_{ij}$ is responsible for breaking rotational symmetry and giving a finite value to Eq.~\eqref{s}.
The eigenvalues and eigenvectors of the matrix $M(\bs q)$ are given by
\begin{widetext}
\begin{align}
 &\hat {\bs n}_L = {1\over q}\left(q_x,q_y,q_z\right)\;\;\;;\;\;\; \l_L(q) = \w_L^2 + c_L^2 q^2\label{eig}\\
 &\hat {\bs n}_{T1} = {{\left(q_y-q_z,q_z-q_x,q_x-q_y\right)\over \sqrt{(q_y-q_z)^2+(q_z-q_x)^2+(q_x-q_y)^2}}}\;\;\;;\;\;\; \l_T(q) = \w_T^2 + c_T^2 q^2\nn\\
 &\hat {\bs n}_{T2} = \hat {\bs n}_{T1} \times \hat {\bs n}_L   \nn
\end{align}
\end{widetext}
This is the exact eigensystem for the case of $\d M = 0$ (or equivalently, $\a = 0$). To compute the eigenbasis for $\a \ne 0$ we treat $\d M$ as a perturbation
\begin{align}
&\d \hat {\bs n}_{T1} = {\hat {\bs n}_{L}^{\mrm T}\cdot \d M(\bs q)\cdot \hat {\bs n}_{T1} \over \l_T(q)-\l_L(q)  }\;\hat {\bs n}_{L} \label{eig_correct}\\
&\d \hat {\bs n}_{T2} = {\hat {\bs n}_{L}^{\mrm T}\cdot \d M(\bs q)\cdot \hat {\bs n}_{T2} \over \l_T(q)-\l_L(q)  }\;\hat {\bs n}_{L}\nn
\end{align}
such that $\hat{\bs e}_{T1,2} = \hat {\bs n}_{T1,2}+\d \hat {\bs n}_{T1,2}+\mc O(\d M^2)$. The perturbative approach can be justified in the limit of
$q\rightarrow 0$ by noting that $\lim_{q\rightarrow 0}\d M(\bs q)=0$ in contrast to $M(\bs q)$ that remains finite, and thus, the eigenbasis \eqref{eig} becomes
exact in this limit. Thus, the perturbation theory is valid in the limit $q^2 \ll (\l_L-\l_T) / \a\approx \w_L^2 / \a \approx 0.7 \left(2\pi /a \right)^2$.

A direct computation using Eq.~\eqref{eig_correct} gives
\be
s\equiv \int {d\W\over 4\pi} \left(\hat {\bs q}\cdot \hat{\bs e}_{T1}\right)^2 =  { r \a^2 q^4 \over 4\pi\left[\l_L(q)-\l_T(q)\right]^2 } \approx {r\a^2 q^4
\over 4\pi \w_L^4 }\label{s_correct}
\ee
where $r\approx 0.239\ldots$ can be expressed as an integral over a lengthy expression of trigonometric functions.
Note that in the last line we assumed $\w_L^2 \gg c_T^2 q^2,\,c_L^2 q^2,\,\w_T^2$.

Now lets make some estimates. In SrTiO$_3$ the longitudinal phonon frequency should be $\w_L = 100\, \mrm{meV}$. We can overestimate the parameter $\a$ by taking
it to be $\a = c_T^2$, where the velocity of the transverse mode is $c_T  \sim 5\, \mrm{meV \,nm}$ (this implies that the dispersion is very anisotropic).
Finally, to get a number we estimate this average at $q = 2k_F$, where $k_F = (3 \pi^2 n)^{1/3}$. At a density $n = 10^{20} \,\mrm{cm^{-3}}$ we find that ${\a
q^2 \over \w_L^2 }\sim 0.01$, and therefore the overlap squared averages to $2\times 10^{-6}$. For $n = 10^{17} \,\mrm{cm^{-3}}$, we get ${\a q^2 \over \w_L^2
}\sim 0.0001$ and therefore the overlap squared is $2\times 10^{-10}$. In contrast W\"olfle and Balatsky estimated $s \approx 0.1$. Clearly, if they take into
account this correction the pairing they found in the s-wave channel will become negligibly small.

We emphasize that the $q^4$ dependence in Eq.~\eqref{s_correct} holds only for modes that create a dipolar excitation within the unit cell. It is this dipolar
coupling that forces the displacement vector to be nearly  perpendicular to $\bs q$. As an example, consider the 6 meV TO mode that is associated with the cubic
to tetragonal transition at 100K. This mode originate as a zone corner mode in the cubic phase that is folded to the zone center and  produces a quadrupole
moment rather than dipole in the unit cell. In this case the factor $s$ can be finite in the limit of small $q$. A reasonably large $\l$ can be obtained by
exchanging this mode at intermediate and high density~\cite{Appel1969} and may supplement the polar phonon mechanism~\cite{gurevich} and contribute to $T_c$,
even though $\l$ will still be small at low density due to the small density of states.
Indeed, there is evidence that this mode contributes to the transport scattering rate around 50K.~\cite{stemmer}

\section{A general formulation of the deformation potential}
Next we consider the general problem of the coupling to a TO mode in a crystal with multiple ions in the unit cell. %To estimate the electron-phonon interaction
we first consider the general potential caused by a deformation.
We make the rigid ionic potential approximation, i.e.,
we write the  potential induced on the electrons by the deformation of the set of lattice deformations $\{\bs U_{j\a} \}$ as
\begin{widetext}
\begin{align}
V_{el-ph}(\bs r)=  \sum_{j\a} \left[V_\a(\bs r-\bs R_{j\a}-\bs U_{j\a})-V(\bs r-\bs R_{j\a}) \right]\approx  \sum_{j\a} \bs U_{j\a} \cdot \nabla V_\a(\bs r-\bs
R_{j\a} )\label{el-ph0}
\end{align}
where $\bs R_{j\a} = \bs R_{j}+\bs \tau_{\a} $ is the position of the $\a$ ion at unit cell $j$ and $V_{\a}(\bs r)$ is the potential it induces. The precise form
of this potential will be discussed later in more detail.
Thus, the matrix element for transitions between electronic states due to this potential is thus given by
\begin{align}
\langle \bs k;n | V_{el-ph}(\bs r) | \bs k' ;n'\rangle &= \sum_{j}\int d^3 r\; \psi_{\bs k n} ^* (\bs r) \bs U_{j\a} \cdot \nabla V_\a(\bs r-\bs R_{j\a}
)\psi_{\bs k' n'} (\bs r)\\
&={i\over \W^2} \sum_{\bs q  }\int d^3 r \;\bs q \cdot\left( \sum_j \bs U_{j\a} e^{i \bs q \cdot \bs R_{j\a}} \right) V(\bs q)e^{i (\bs q+\bs k-\bs k') \cdot \bs
r}\chi_{\bs k n}(\bs r)\chi_{\bs k' n'}^*(\bs r)\nn\\
&={i\over \W^{3/2}} \sum_l e^{i (\bs q+\bs k - \bs k')\bs R_l}\sum_{\bs q}{1\over v_{uc}}\int_{uc} d^3 r \;\bs q \cdot \bs U_{\bs q \a} V_\a({\bs q}) e^{i (\bs
q+\bs k-\bs k') \cdot \bs r}\chi_{\bs k n}(\bs r)\chi_{\bs k' n'}^*(\bs r)\nn\\
&={i\over \W^{3/2}  } \sum_{\bs G }V_\a({\bs Q})\;\bs Q \cdot \bs U_{\bs k-\bs k'\a}\int_{uc}{d^3 r\over v_{uc}} \;\bs e^{i \bs G \cdot \bs r}\chi_{\bs k n}(\bs
r)\chi_{\bs k' n'}^*(\bs r)\nn
\end{align}
Here $\W = N a^3$ is the total volume, $a$ is the lattice constant, $\bs Q = \bs k'-\bs k+\bs G$ represents the conservation of crystal momentum, the $\psi_{\bs
k n}(\bs r) = {e^{i\bs k \cdot \bs r}\over \sqrt{\W}} \chi_{\bs k n}(\bs r) $ are the Bloch wave functions of an electron in band $n$ and the integral is over
the unit-cell.
Note that $\bs U_{\bs q,\a}=\bs U_{\bs q+\bs G,\a}$. Thus, we can define the transition matrix element
\be
M_{nn'}^{\bs G\a}(\bs k, \bs k')\equiv {iV_\a({\bs Q})\over  v_{uc}}  \int_{uc}d^3 r e^{i \bs G \cdot \bs r}\chi_{\bs k n}(\bs r)\chi_{\bs k' n'}^*(\bs r)
\ee
such that the electron-phonon coupling is given by
\be
H_{el-ph} = {1\over \W^{3/2}}\sum_{\bs G \a \bs k \bs k'}M_{nn'}^{\bs G\a}({\bs k, \bs k'})(\bs k' - \bs k +\bs G)\cdot \bs U_{\bs k-\bs k'\a}c^\dag_{\bs k n}
c_{\bs k' n'} \label{el-ph-full}
\ee
This expression is quite general. Notice that there are terms proportional to $\bs G$, which allow a finite coupling in the limit $q=|\bs k - \bs k'|\rightarrow
0$. We comment on these terms
%in a footnote~\footnote{Regarding the terms with finite $\bs G$ in the limit of zero momentum transfer $\bs k = \bs k'$. We note that when both time-reversal
and inversion symmetry are present, one can show that the sum of two contributions, $\bs G$ and $-\bs G$, cancel each other if also $n = n'$ (i.e. intraband
scattering). That is, $M_{nn}^{\bs G}(\bs k,\bs k) = M_{nn}^{-\bs G}(\bs k,\bs k)$. To see this we note that with TRS we have $\chi_{\bs k n}^*(\bs r)=\chi_{-\bs
k n}(\bs r)$ and with inversion we have $\chi_{\bs k n}(\bs r)=\chi_{-\bs k n}(-\bs r)$. Thus, the product $\chi_{\bs k n}^*(\bs r) \chi_{\bs k n} (\bs r)$ is an
even function of $\bs r$, and thus its Fourier series is an even function of $\bs G$. Thus, the conclusion, is that with these symmetries present the electron
coupling to transverse optical phonons at zero momentum transfer must include interband processes.
%},
at the end of this section, but since the authors of Ref.~\onlinecite{Wolfle2018} did not consider the finite $\bs G$ terms we continue to follow their analysis
and consider only normal processes with $\bs G = 0$. Additionally, they also only considered a single band approximation, therefore they arrive at a result of
the form
\begin{align}
H_{el-ph}&= i\sum_{\bs q\a } V_\a({\bs q})\bs q \cdot \bs U_{\bs q \a} \rho_{-\bs q}\label{H_e-ph}
\end{align}

Let us write $ V_\a({\bs q})=Z_\a V_C({\bs q})+V_{ps,\a}(\bs q)$ where
\be
V_C(\bs q)= {4\pi e^2 \over \ve_{\infty } q^2 }
\ee
is the Coulomb interaction, $Z_\a$ is the charge on ion $\a$ such that charge neutrality gives $\sum_\a Z_\a = 0$ and $V_{ps,\a}(\bs q)$ is what is left over,
which we refer to as the pseudo-potential of ion $\a$.
Then Eq.~\eqref{H_e-ph} becomes
\be
H_{el-ph}= i\sum_{\bs q } V({\bs q})\bs q \cdot \bs u_{\bs q } \rho_{-\bs q}+i\sum_{\bs q\a } V_{ps, \a}({\bs q})\bs q \cdot \bs U_{\bs q \a} \rho_{-\bs q}
\label{H_e-ph2}
\ee
\end{widetext}
where $\bs u_{\bs q}=\sum_{\a } Z_\a \bs U_\a({\bs q})$ is the displacement, which is proportional to the unit-cell dipole and corresponds to the soft phonon
discussed in Section \ref{Sec:polariaztions}. This discussion there applies: in particular, the average of
$(\hat {\bs q}\cdot \hat{\bs u}_{\bs q})^2 $
is proportional to $q^4$ as given by Eq.~\eqref{s_correct}.
As a result, the contribution from the Coulomb interaction, the only one kept by W\"olfle and Balatsky, is entirely negligible.
On the other hand, the second term in Eq.~\eqref{H_e-ph2} is proportional to $\bs u'\cdot\bs q$ where $\bs u_{\bs q}'= \sum_{\a } V_{ps, \a}({\bs q}) \bs U_{\bs
q \a}$ and where $V_{ps, \a}$ is the short-ranged pseudo potential of ion $\a$.
$\bs u'$ is in general not parallel to $\bs u$ and does not need to be perpendicular to $\bs q$ even in the $\bs q\rightarrow 0$ limit. However, for cubic and
tetragonal crystal structure, as in SrTiO$_3$, symmetry constraints all the individual displacements $\bs U_{\bs q \a}$ to be colinear near the zone center. This
is clearly the case for $\bs q$ along a symmetry direction such as the $x$ axis and it is easy to see that it continues to hold for arbitrary direction because
the force matrix is non-singular in the $\bs q$ goes to zero limit. Thus, $\bs u$ is parallel to all $\bs U_{\bs q \a}$ and we may conclude that also this term
is suppressed by the same factor of $s\propto q^4$ and is therefore negligible.

To conclude, we have shown that the gradient coupling of the soft TO mode to electronic density in SrTiO$_3$ is dramatically suppressed near the zone center.
This is due to long-ranged Coulomb forces, which bend the polarization to become truly transverse to $\bs q$. The result holds both for the long-ranged Coulomb
repulsion and the short range pseudo potential. We find confirmation to our results in a recent {\it ab initio} calculation of the electron-phonon coupling in STO~\cite{Zhou2018}, where the coupling to the TO phonon is found to be weak and decreases at small $q$.
%It is also important to point out that we have already discussed this issue in the appendix of Ref.~\cite{Ruhman2016}, there exists a vector coupling enabled by spin-orbit coupling. Despite being much
%greater than the gradient coupling, the former is also expected to be weak.
% The proper calculation of the average of $(\hat {\bs q}\cdot \hat {\bs U'}(\bs q))^2 $  will require a detailed modelling but one should keep in mind that it
%is subject to the constraint  that $(\hat {\bs q}\cdot \hat{\bs U}(\bs q))^2 $ is small. Furthermore, $V_{ps,\a}(\bs q)$ is generally considerably smaller than
%the Coulomb repulsion and certainly smaller than the value of 35 eV used by Wolfle and Balatsky. Thus while the second term in Eq.~\eqref{H_e-ph2} does
%contribute, its value should be much smaller than the  estimate by Wolfle and Balatsky. We have direct evidence for this from transport data as discussed in the
%next section.

Regarding the terms with finite $\bs G$ in Eq.~\eqref{el-ph-full} we note that in the limit of zero momentum transfer $\bs k = \bs k'$ and when both
time-reversal and inversion symmetry are present, one can show that the sum of two contributions, $\bs G$ and $-\bs G$, cancel each other if  $n = n'$ (i.e.
intraband scattering). That is, $M_{nn}^{\bs G}(\bs k,\bs k) = M_{nn}^{-\bs G}(\bs k,\bs k)$. To see this we note that with time reversal symmetry we have
$\chi_{\bs k n}^*(\bs r)=\chi_{-\bs k n}(\bs r)$ and with inversion we have $\chi_{\bs k n}(\bs r)=\chi_{-\bs k n}(-\bs r)$. Thus, the product $\chi_{\bs k
n}^*(\bs r) \chi_{\bs k n} (\bs r)$ is an even function of $\bs r$, and thus its Fourier series is an even function of $\bs G$.  Concerning interband scattering,
for states with $\bs k$ near the zone center, they are either even or odd under inversion. interband scattering between even and odd states is allowed for states
near the zone center. However, in STO, the states of interest near the Fermi level are d wave which are even. Therefore we conclude that the finite $\bs G$
processes are negligible in STO at low density.

\section{Bounding the coupling using transport data}
In this section we roughly estimate the coupling to the TO mode based on the resistivity measurements in Ref.~\onlinecite{lin2015scalable}. If a coupling of the form
Eq.~\eqref{el-ph} exists we anticipate the phonon limited transport lifetime at temperatures higher than the mode frequency
\be
{\hbar \over \tau} \approx 2\pi k_B T \l
\ee
where $\l$ is the BCS coupling strength. It is important however to note that the formula above applies to a flat phonon band. For the soft TO mode this applies
only when $k_F < \w_T / v_s \approx 3 \times 10^{6} cm^{-1}$, which corresponds to a density of $10^{18}$ cm$^{-3}$. At this density Ref.~\cite{lin2015scalable}
measure an inverse lifetime $\tau_{tr}^{-1} \approx 0.5$ ps$^{-1}$ at a temperature of $T = 30$. Estimating the coupling we get that $\l\leq {\hbar /\tau \over
\pi k_B T } \approx 0.04$ which is in strong disagreement with the results of Ref. ~\onlinecite{Wolfle2018}, which find $\l \sim 0.2$. For higher densities, we
can add a factor $ (\w_T/ v_s k_F)^2$ to  the RHS of Eq (26) to keep scattering by only the modes that are almost flat. For a density of  $5 \times  10^{18}$
cm$^{-3}$, we still get a bound of $\l \sim 0.15$, giving a $T_c$ of about  $10^{-3} \w_{TO}$ according to BCS theory.

\section{The inclusion of the high-frequency LO mode in the Eliashberg framework }
In addition to the soft TO mode, which was discussed in this comment in great detail, the authors of Ref.~\onlinecite{Wolfle2018} have also considered the
dynamics of the screened Coulomb potential as a pairing mechanism. In particular, they considered the dynamics associated with the high-frequency LO mode. They
argue that even when the phonon frequency is greater than the Fermi energy, the Eliashberg still captures the essential physics up to the cut-off $\w_c$, which is determined by the criterion that the quasi-particles scattering rate is small compared to frequency. We disagree with this and point out that the Eliashberg equation is based on the approximation of keeping the ladder diagrams with the leading logarithmic divergence. For frequencies above the Fermi energy, the ladder diagrams no longer carry the logarithm and many other diagrams contribute equally (for example see the diagrams in Fig.2 of Ref.\,\onlinecite{PhysRevB.29.6132}). Whether the quasi-particles are well defined or not, is not the only issue.

Historically, Takada~\cite{Takada1980} treated the dynamically screened interaction in the same way and integrated the Eliashberg equation up to very high
frequencies, to obtain a relatively large $T_c$, However,  this procedure has been criticized by showing that vertex corrections are
large~\cite{PhysRevB.29.6132}. In  Ref.~\onlinecite{Wolfle2018}, the consequence of taking a large cut-off is that  they found a $T_c$ dome with $T_c$ as large
as 0.45 K even for densities as low as $10^{17} cm^{-3}$, in clear disagreement with experiment (see the green curve in their Fig. 2).

In Appendix A.3  the authors of Ref.~\onlinecite{Wolfle2018} attempted to justify their procedure by arguing  that the vertex correction is small because the
coupling  is weak. However, they used a full screened interaction, taking the static limit of screening not only for the electrons, but also for the phonons in
their eq.A34. We note that according to ref \onlinecite{gurevich} the attraction comes precisely from the frequency dependence of the phonon part of the
dielectric function
$\e_{ph}(\w)= \e_{\infty} (\w_{LO}^2 +  \w_n^2 )/ (\w_{TO}^2 +  \w_n^2)$ where $\w_n$ is the Matsubara frequency. The attraction comes from the difference in
potential with the dielectric function evaluated  between the frequencies $\w_{LO}$ and $\w_{TO}$. This difference  is much larger than the staic limit used in
Ref.~\onlinecite{Wolfle2018}.  By using the static lomit, these authors have under-estimated the couplng by orders of magnitude.   The correct estimate should be
taken at a range of  frequencies  up to the LO mode and  the vertex correction is non-negligible. Thus, there is no reason why only the ladder diagrams can be
kept and the results of using the Eliashberg equation up to high cut-off are generically inaccurate.

\section{Acknowledgments}
We thank Peter W\"olfle for pointing out Ref.~\onlinecite{Zhou2018}.
JR acknowledges work to appear with Zhen Bi and Vladyslav Kozii which is closely related to some of the results presented here. PAL acknowledges the support of
DOE  under grant no. FG02-03-ER46076.

\end{document}